\documentclass[11pt]{article}
\pdfoutput=1
\usepackage[utf8x]{inputenc}
\usepackage{amssymb,amsmath,mathrsfs,enumerate}
\usepackage{rotate,multicol}
\usepackage{graphicx,wrapfig}
\usepackage{tikz,graphics,color,fullpage,float,epsf,subcaption}
\usepackage{pdfpages}
\usepackage{mathtools}
\usepackage{multirow}
\usepackage[textfont=it,labelfont=bf]{caption}
\usepackage[colorlinks=true,
  linkcolor=red,
  urlcolor=blue,
  citecolor=blue]{hyperref}
\usepackage{color}
\usepackage{soul}
\usepackage{cite}
\usepackage{verbatim}
\usepackage{array}
\usepackage[compat=1.0.0]{tikz-feynman}
\usepackage[symbol]{footmisc}
\long\def\rpl#1!!#2!!{\textcolor{red}{#1} \textcolor{blue}{#2}}

\newcommand{\fulleqref}[1]{Equation~\!\eqref{#1}}
\newcommand{\fullfigref}[1]{Figure~\!\ref{#1}}

\newcolumntype{C}[1]{>{\centering\arraybackslash}p{#1}}

\newmuskip\pFqskip
\mathchardef\pFcomma=\mathcode`, 

\evensidemargin=\oddsidemargin
\allowdisplaybreaks

\begin{document}

\begin{center}
{\Large \bf  Leptophilic ALPs with TWIST data for polarized muon decays } \\
  \vspace*{1cm} {\sf Ankita Budhraja~$^{a,}$\footnote[1]{abudhraj@nikhef.nl},~Samadrita 
    Mukherjee~$^{b,}$\footnote[2]{samadritam@iisc.ac.in},~Sahana
    Narasimha~$^{c,d,}$\footnote[3]{sahana.narasimha@oeaw.ac.at}} \\
  \vspace{10pt} {\small \em 
    $^a$Nikhef, Theory Group, Science Park 105, 1098 XG, Amsterdam, The Netherlands \\
    $^b$Center for High Energy Physics, Indian Institute of Science, Bengaluru 560012, India,\\
    $^c$Institut für Hochenergiephysik, Österreichische Akademie der Wissenschaften, \\
Nikolsdorfer Gasse 18, A-1050 Wien, Austria,\\
    $^d$University of Vienna, Faculty of Physics, Boltzmanngasse 5, A-1090 Wien, Austria.\\
  }
  
  \normalsize
\end{center}

\renewcommand*{\thefootnote}{\arabic{footnote}}
\setcounter{footnote}{0} 
\begin{abstract}
 We study the production of axion-like particles (ALPs) in association with electrons and neutrinos in the muon decay process. For this purpose, we compute the decay width of the muon to a four-body channel using a $d=7$ effective operator that couples the ALP to the Standard model fermions, namely leptons and neutrinos. Assuming a dominant coupling of the ALP to the dark sector, we only consider ALP decays to invisible final states. To obtain constraints on our model using the existing measurements, we leverage data from the TRIUMF Weak Interaction Symmetry Test (TWIST) experiment and obtain bounds on the ALP-lepton coupling for masses in the range of $0 < m_{\phi} < m_{\mu}/4$, as allowed by kinematics. Using the precision of current TWIST measurements, we obtain an order of magnitude estimation necessary for future searches to further constrain the parameter space for such a setup. Furthermore, we find that keeping realistic considerations, the new physics contribution can possibly be enhanced even with a minimalistic modification to the fiducial area used in the experiment potentially allowing for stringer constraints. At the end, in an attempt to relax the assumption that ALP decays to invisible only, we also investigate its stability and find potential longevity within collider environments for the mass range considered in this study.
\end{abstract}


\section{Introduction}
Over the past few decades, precise measurements of electroweak observables at high-energy colliders, along with focused investigations in flavor physics and neutrino experiments, have demonstrated striking alignment with the Standard Model (SM). While direct searches at the colliders effectively restrict new physics interactions with quarks or gluons, significant parameter space remains open for models featuring light new particles solely interacting via electroweak (EW) forces. One such well-motivated extension of the SM are the {\it axions}~\cite{Weinberg:1977ma, Wilczek:1977pj} and {\it axion-like particles}  (ALPs)~\cite{Kim:2008hd, Marsh:2015xka, Brivio:2017ije}. Within the SM framework, the spontaneous breaking of chiral symmetries leads to pseudo-Nambu-Goldstone bosons, exhibiting ALPs characterized by a shift symmetry $a \simeq a+ 2\pi f_a$, where $f_a$ represents the axion decay constant. Although quantum chromodynamics (QCD) breaks this shift symmetry for typical axions, resulting in their mass, this effect can be decoupled from the QCD scale for light axion masses.
Beyond the original motivation to
address the strong CP problem, ALPs are one of the primary targets for the large-scale experimental effort to search for
hidden sectors \cite{Jaeckel:2010ni, Ringwald:2012hr, Essig:2013lka}. ALPs are primarily defined by their interactions, which are linked through derivatives and enjoy a fundamental global symmetry at high energies referred to as Peccei-Quinn symmetry \cite{Peccei:1977hh, Peccei:1977ur}. 
The charges of the SM fields under this global symmetry determine the specific interactions of ALPs to SM particles.

In this article, we direct our attention to ALPs primarily connected with leptons. 
Detection of such “leptophilic” ALPs has been widely
contemplated within terrestrial experiments \cite{Konaka:1986cb, Riordan:1987aw, Bjorken:1988as, Bross:1989mp, Scherdin:1991xy, Tsai:1989vw, Bassompierre:1995kz, Izaguirre:2016dfi, Marciano:2016yhf, Berlin:2018bsc, AristizabalSierra:2020rom, Gori:2020xvq, Bauer:2020jbp, Bauer:2021mvw}. Additionally, the potential for ALP detection through charged meson decays (e.g., $\pi^+$, $K^+$) and W boson decays has been explored \cite{Altmannshofer:2022ckw, Lu:2022zbe, Lu:2023ryd}. Here, our focus shifts to investigating the production of light ALPs which we denote as $\phi$ alongside electrons and neutrinos in the Standard Model muon decay process, $\mu^{\pm} \to e^{\pm} \nu_{\mu} \bar{\nu}_{e} \phi$. This process stands out as the cleanest production channel due to (a) the definite measurement of muon lifetime, and (b) the absence of hadrons in the decay process, enabling a cleaner search strategy with reduced uncertainties associated with hadronic in-states.

Owing to the precise measurement of the muon lifetime, one can hope to obtain stringent constraints on the strength of the new physics (NP) couplings via the precision of these measurements. However, it is hindered by the fact that the observed value may not necessarily be identical to the SM prediction as the latter relies on a precise knowledge of the input parameters that are determined through a global fit of various experimental data.  In an attempt to reduce these uncertainties, one possibility would be to utilize differential measurements.
For this purpose, we will utilize the existing results of the TRIUMF Weak Interaction Symmetry Test (TWIST) experiment \cite{TWISTprd2011} to understand the precision of these measurements and the potential for detecting ALPs produced primarily in association with leptons. At this point, it should be further noted that in order to accurately evaluate the constraints on ALP couplings, it is essential to specify not only the channel through which ALPs are produced but also their decay mechanism. In this context, we assume that ALP remains stable within colliders and does not decay into observable SM particles. Such a scenario can be achieved, for example, by assuming a sufficiently strong coupling of the ALP to a stable dark sector~\cite{Nomura:2008ru, Bharucha:2022lty, Armando:2023zwz, Buttazzo:2020vfs}.

The Fermi theory of muon decay is a simple example of an effective field theory (EFT) where the intermediate W-boson being heavy can be integrated out for the low-energy processes of interest. Treating the muon decay process in an effective framework, at the scale of $m_{\mu}$ (the muon mass), the only relevant degrees of freedom in the SM are electrons, neutrinos, and photons. To be specific, the leading operator that gives rise to muon decay to electrons and neutrinos is given by
\begin{equation}
 \mathcal{L}\supset G_{F} \left(\overline{\mu} \gamma^{\rho}P_L \nu_{\mu}\right)\left( \bar{\nu}_{e} \gamma_{\rho}P_L e  \right) , 
 \label{eq:SM-lagrangian}
\end{equation}
with $G_F$ being the fermi constant and $P_L = (1-\gamma_5)/2$ is the projection operator. We extend the above EFT to include the beyond standard model (BSM) effects by introducing an additional real (light) scalar particle $\phi$ and take the BSM operator in the context of muon decay as,
\begin{equation}\label{new_phi_operator}
\mathcal{L}_{B S M} \supset \frac{g_{\phi l}}{m_l} G_{F} \phi \,\left(\overline{\mu} \gamma^{\rho}P_L \nu_{\mu}\right) \left(\bar{\nu}_{e} \gamma_{\rho}P_L e\right). 
\end{equation}
Here, $g_{\phi l}$ is a dimensionless complex coupling constant and we have chosen the dimensionful normalization scale, $m_l$, as the mass of the charged lepton, $2 m_e$ to be specific.
Note that even though \fulleqref{new_phi_operator} implies a non-zero width
of $\phi$ to SM fermions (when allowed by kinematics), this width would be
phase-space suppressed.  
Consequently,  the fractional width of \(\phi\) to SM particles can be made negligible by assuming a
marginal coupling of $\phi$ to the dark sector.\footnote{We do not further delve into the construction of a concrete model for the dark sector in this paper as our main goal is to study ALP signatures to missing transverse energy.} Since the neutrino is not directly detectable and its transverse momentum is deduced by imposing a transverse momentum conservation, we can, in principle (with the above construction), have the light BSM degree of freedom $\phi$ that goes invisibly contributing to the process as well.

We further argue that even though constraints on the ALP parameter space exist from decays of heavier mesons like $\pi$, $K$ decays; in order to exhaust all the existing measurements, it is indeed necessary to obtain complementary bounds from the much cleaner leptonic environments as well. We find that the studies on ALP parameter space from meson decays are more constraining than the bounds obtained in our work. However, it is instructive to point out that in these studies, it is not possible to disentangle the ALP-lepton coupling from the ALP-quark coupling as discussed in Ref.~\cite{Bandyopadhyay:2021wbb}. Therefore, the channel that we consider not only provides for a cleaner environment but also allows us to completely isolate the effect of ALP-lepton couplings on the decay spectrum.  

The article is organized as follows. In Sec.~\ref{sec:three-body}, we briefly outline the well-studied differential distribution of muon decay within the SM and provide some details of the TWIST experiment that we utilize. In Sec.~\ref{sec:four-body}, we study the same process including the new decay mode containing an additional scalar contributing to missing energy in the detector. We also study the dependence of this process to the mass of the new scalar. We then provide the details of our simulation and analysis in Sec~\ref{sec:simulation_details}. Finally, we extract the bound on the effective coupling of our NP interaction by utilizing the TWIST differential data, in Sec.~\ref{sec:constraints}. Here we find that the heavier the mass of the scalar is, the looser the bound on its effective scale of interaction becomes. We present a brief summary and conclude in Sec.~\ref{sec:conclusions}. We have also moved some specific details to the appendices. In Appendix~\ref{appendix-A}, we provide the detailed form of the polynomial that appears for the four-body decay spectrum as a function of the mass of the light scalar. In Appendix~\ref{appendix-B}, we provide speculations on the possible completions for the proposed effective operator.

\section{Polarized muon decay in the SM and the TWIST experiment}
\label{sec:three-body}
In the SM, the muon decays primarily through weak interactions mediated by the W-boson. As the mass of the W-boson mediating the process is much larger than the muon mass, one can study the process equivalently using the effective operator introduced in \fulleqref{eq:SM-lagrangian}. 
The differential decay distribution for a polarized $(\mu^{\pm}\to e^{\pm}\nu \bar{\nu})$ muon(antimuon),  
decaying in its rest frame, within SM is given as \cite{Michel50},  
\begin{align}
         \frac{d^{2} \Gamma_3(\mu^{\pm}\to e^{\pm}\nu \bar{\nu})}{d E_e\, d\left(\cos \theta_{e}\right)} & = \frac{G_{F}^{2} m_{\mu}^{2}}{24 \pi^{3}} \vert\vec{k}_e\vert \left\{3 E_e-\frac{4 E_e^{2}}{m_\mu}+\frac{3\, m_e^2}{m_\mu^2} E_e  
          -\frac{2\,m_{e}^{2}}{m_\mu} \pm \vert\vec{k}_e\vert P_{\mu} \cos \theta\,\left(\!\!1 -\frac{4 E_e}{m_\mu}\!
        -\!\frac{3\, m_e^2}{m_\mu^2} \right)\right\} ,
        \label{3-body-decay}
   \end{align}
    with $\vert\vec{k}_e\vert = \sqrt{E_e^{2}-m_{e}^{2}}$. Here, $E_e$ is the energy of the outgoing electron (positron) being observed, and $\theta_e$ is the angle of the electron (positron) to the polarization axis of the muon (antimuon) beam with polarization degree $P_\mu$. From the simple three-body kinematics, in the muon's rest frame, it is straightforward to see that $m_{\mu}/2$ is the maximum allowed energy of the emitted electron (positron). Any additional BSM scalar associated with this process would therefore : (a) be phase space suppressed and (b) shift the energy spectrum of the observed electron (positron) coming from the decay. We will discuss the details of such a setup in the next section. Here we first describe the experimental setup that we utilize.

The TRIUMF Weak Interaction Symmetry Test (TWIST) experiment situated in Canada \cite{TWIST_Apparatus:2005, TWIST:2011egd, TWIST:2011aa, TWIST:2008myj, TWIST:2004lzd, TWIST:2004hce}
observes the momentum-angle spectrum of the electrons (positrons) emitted from a linearly polarized muon (antimuon) decaying at rest.
Muons are produced as follows: a $500$ MeV proton beam 
collides a carbon (graphite) target producing a large number of positively charged pions, $\pi^+$.
A subset of these $\pi^+$ particles comes to a halt within the target material, undergoing decay at rest, predominantly giving rise to positive muons ($\mu^+$) and neutrinos.  
Positive muons are preferred for this study due to the propensity of negative muons to be captured by the Coulomb field of the target nuclei, which would distort the decay spectra compared to free muons. The experiment achieves a muon stopping rate ranging from 2000 to 5000 muons per second, and an impressive amount of $10^{10}$ events were recorded. This experimental analysis then allows for precision studies of the electron spectrum emitted during the muon decay, enabling the test of the $V-A$ structure of weak interactions. The minimal theoretical uncertainties associated in this experiment additionally make this search channel particularly useful for exploring light and/or weakly coupled BSM phenomena. 

We now describe the four-body decay and its kinematics, provide the details of our simulation and finally utilize the data obtained from the TWIST collaboration for the differential decay spectrum to obtain the constraints on the ALP-lepton coupling from the existing measurement.~\footnote{We thank Arthur Olin for providing us the experimental data as well as their simulation fits which we could utilize to obtain an estimate for the effect of non-inclusion of higher order QED radiations as well as the detector effects into our clean theoretical setup.}

\section{New physics modified polarized muon decay }
\label{sec:four-body}

 With the inclusion of the effective NP operator as depicted in \fulleqref{new_phi_operator}, we compute the differential decay spectrum for the muon decaying to an additional light scalar in association with the usual final states.
This new light scalar $\phi$ goes invisible contributing to the total missing transverse energy in the detector. In the limit $m_{\mu} > m_{\phi} \sim m_e$, the NP decay mode,
\begin{equation*}
\mu^{-}(p_{\mu}) \to e^{-} (k_e) + \nu_{\mu} (k_{\nu_{\mu}}) + \bar{\nu}_e(k_{\bar{\nu}_e}) + \phi (k_{\phi})\, ,
\end{equation*}
is kinematically allowed. 
As the ALP only has a pseudo-scalar coupling to SM fermions, the square of the transition amplitude for the above four-body decay of muon is essentially the same as in the case of a standard
polarized muon decay with an additional factor of $g_{\phi l} / m_l$, accounting for the NP coupling.
 \begin{figure}[!htb]
\begin{center}
\begin{tikzpicture}
  \begin{feynman}
    \vertex (a) {\(\mu^{-}\)};
    \vertex [right=of a] (b);
    \vertex [above right=of b] (f1) {\(\nu_{\mu}\)};
    \vertex [below right=of b] (c);
    \vertex [above right=of c] (f2) {\(\overline \nu_{e}\)};
    \vertex [below right=of c] (f3) {\(e^{-}\)};
    
    \diagram* {
      (a) -- [fermion] (b),
      (b) -- [fermion] (f1),
      (b) -- [anti fermion] (f2),
      (b) -- [fermion] (f3),
    };
  \end{feynman}
  
  \node[draw, circle, fill=black, inner sep=2pt] at (b) {};
\end{tikzpicture}
\hspace{3cm}
\begin{tikzpicture}
  \begin{feynman}
    \vertex (a) {\(\mu^{-}\)};
    \vertex [right=2.5 cm of a] (b);
    \vertex [above right=of b] (f1) {\(\nu_{\mu}\)};
    \vertex [right=2 cm of b] (f2) {\(\overline \nu_{e}\)};
    \vertex [below right= 2 cm of b] (f3) {\(\phi\)};
     \vertex [below = 1.5 cm of b] (f4) {\(e^{-}\)};
  \diagram{
      (a) -- [fermion] (b) -- [fermion] (f1),
      (b) -- [anti fermion] (f2),
      (b) -- [blue, very thick, scalar] (f3),
      (b) -- [fermion] (f4),
    };
  \end{feynman}
  \node[draw, circle, fill=black, inner sep=2pt] at (b) {};
\end{tikzpicture}
\caption{Feynman Diagrams for muon decay processes in SM (left) and with an extra light scalar (right).}
\end{center}
 \end{figure}
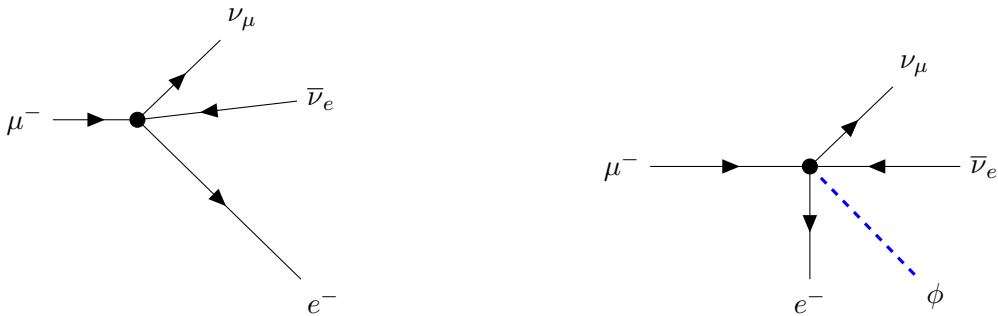
The decay amplitude consists of four phase-space integrations for each of the final states. The integrations over the neutrino momenta are carried out trivially. We also integrate out the azimuth dependence on angles $\Phi_e$ and $\Phi_{\phi}$ in the integrals from $0$ to $2\pi$. As the experiment is only sensitive to the momentum and the angle of positron being emitted with respect to the muon polarization axis, we carry out the integrals containing dependence on $E_{\phi}$ as well, leaving the $E_e$ and $\cos\theta_e$ as the variables in the decay spectrum. The differential distribution of this four-body decay at this stage reads as,
\begin{align}
\frac{d^{2} \Gamma_4(\mu^{\pm}\to e^{\pm}\nu \bar{\nu} \phi)}{d E_{e} d \cos \theta_{e}} &=\frac{G_{F}^{2} g_{\phi l}^2}{192 \pi^{5} m_l^2}\vert\vec{k}_e\vert \times\bigg\{\Big( 6 m_{\mu}^{2} E_{e}+6 m_{e}^{2} E_{e} 
+2 m_{\phi}^{2} E_{e}-8 m_{\mu} E_{e}^{2}-4 m_{e}^{2} m_{\mu} \nonumber\\
& \mp 2 m_{\mu}^{2}\vert\vec{k}_e\vert P_{\mu} \cos \theta_{e}  
 \mp 6 m_{e}^{2}\vert\vec{k}_e\vert P_{\mu} \cos \theta_{e}  
 \mp 2 m_{\phi}^{2}\vert\vec{k}_e\vert P_{\mu} \cos \theta_{e} \nonumber \\
&  \pm 8 m_{\mu} E_{e}\vert\vec{k}_e\vert P_{\mu} \cos \theta_{e} \Big) I_{1} 
+\Big(4 m_{e}^{2}+8 E_{e}^{2}  
-12 m_{\mu} E_{e} \pm 4 m_{\mu}\vert\vec{k}_e\vert P_{\mu} \cos \theta_{e}  \nonumber \\
& \mp 8 E_{e}\vert\vec{k}_e\vert P_{\mu} \cos \theta_{e} \Big) I_{2} 
+4 E_{e} I_{3} 
 \pm \frac{4}{3}\vert\vec{k}_e\vert P_{\mu} \cos \theta_{e} I_{4}\bigg\} \, ,
\label{4-body-decay}
\end{align}
%
where, for all practical purposes, we choose the magnitude of the polarization vector of the muon, $P_{\mu}$, to be a unit vector along the $\hat{z}$ direction. The $I_{1,2,3,4}$ are different integrals in $E_{\phi}$ which have the form,
\begin{eqnarray*}
I_{1}&=&\displaystyle \int \sqrt{E_{\phi}^{2}-m_{\phi}^{2}}\, d E_{\phi}\, , \\
I_{2}&=&\displaystyle \int E_{\phi} \sqrt{E_{\phi}^{2}-m_{\phi}^{2}}\, d E_{\phi}\, , \\
I_{3}&=&\displaystyle \int E_{\phi}^{2} \sqrt{E_{\phi}^{2}-m_{\phi}^{2}}\, d E_{\phi}\, , \\
I_{4}&=&\displaystyle \int\left(E_{\phi}^{2}-m_{\phi}^{2}\right)^{\frac{3}{2}}\, d E_{\phi}\, .
\end{eqnarray*}
%
These integrals are bounded by kinematic constraints on the four-body decay spectrum. The lower limit being the rest mass of the scalar particle, $m_{\phi}$, while for the upper limit, we utilize the four-momentum conservation which yields 
\begin{equation}
         m_{\phi} \leq  E_{\phi} \leq \frac{m_{\mu}}{2} + \frac{m^2_e}{2 m_{\mu}} + \frac{m^2_{\phi}}{2 m_{\mu}} - E_e\, ,
          \label{eq:integral-limits}
    \end{equation}
with $E_e$ being the energy of the emitted electron.

From the expression of the differential distributions given in Eqs.~\ref{3-body-decay} and ~\ref{4-body-decay}, it is also straightforward to obtain the corresponding total decay widths for the SM and BSM scenarios.
In \fulleqref{3-body-decay}, integrating over electron energy and $\cos\theta_e$ produces the well-known total muon decay width, whereas, from \fulleqref{4-body-decay}, we obtain the total 4-body decay width as a function of the mass of the BSM scalar. 
\begin{align}
\label{eq:Gamma3_Gamma4}
    \Gamma_{3b} &= \frac{G_F^2\, m_\mu^5}{192\, \pi^3} \, , \\ 
    \Gamma_{4b} &= \frac{G_F^2}{192\, \pi^5}\frac{g_{\phi l}^2}{m_l^2} \mathcal{F}(m_{\mu},m_e,m_{\phi})\,  \label{eq:Gamma4_function},
\end{align}
where $\mathcal{F}$ is a polynomial function of the masses involved in the process, the detailed expression of which is given in Appendix \ref{appendix-A}.
In the limiting case where, $m_{\phi} \ll m_{\mu}$, we are left with a simple expression for the total decay width given by,
\begin{equation}
    \Gamma_{4b} = \frac{G_F^2\, m_\mu^5}{192\, \pi^3} \Big(\frac{17\, m_\mu^2}{6720\, \pi^2} - \frac{m_e^2}{96 \pi^2} \Big) \frac{g_{\phi l}^2}{m_l^2}\, .
    \label{Gamma_4}
\end{equation}
In \fulleqref{Gamma_4}, we prefer to show the dependence on the electron mass also.
Having derived the total four-body decay width along with the momentum-angle spectrum for a polarized muon decay process in the context of both
SM and BSM, we will now show where the new predicted decay is expected to show up in experiments. As argued previously, we expect a shift of the observed energy spectrum for the lepton along with a significant phase space suppression by roughly a factor of $16\pi^2$. 
\section{Simulation details}
\label{sec:simulation_details}

In this section, we discuss the details of our simulations. To simulate the decay spectrum, we utilize the differential decay distributions given in Eqs. \ref{3-body-decay} and \ref{4-body-decay} for the SM and the NP scenario. The details of the steps we follow are listed below:

\begin{itemize}

\item We generate pseudo-experiments by allowing the same number of muons to decay as described in the TWIST experiment for the SM decay \cite{TWIST:2011egd, TWIST:2011aa}. We use the same number of bins in the energy and angular variable as used in the experimental simulations, with the probability density in each bin described by the double differential spectrum.

\item  We impose cuts on our generated 2D distribution that correspond to the fiducial regions used in the experiment. The choice of these regions is driven by two conditions. It should be wide enough to increase the number of events used
in the fit of decay parameters but must also exclude regions that may have significant biases. The cuts in the angular variable that are then employed are $|\cos\theta| > 0.54$ and $|\cos\theta| < 0.96$. These are due to the fact that high-angle and low-angle helical
tracks may not be reconstructed reliably. The events with positron momentum $p > 52.0$ MeV are not kinematically allowed and hence excluded for the decay parameter fit. The longitudinal and transverse momentum cuts: $p_l > 14.0$ MeV, $p_t > 10.0$ MeV, and $p_t < 38.0$ MeV, which are used to make sure that we retain positrons within the instrumented regions of the detector and whose radii can be reconstructed reliably. Note that we closely follow these cuts in our pseudo-experiment to mimic the experimental setup as closely as possible.  

\item  We normalize the value of the differential spectra in the selected fiducial regions such that the sum of values in all the bins that fall in each of the fiducial regions comes to 1, respectively.

\item We repeat all the steps above for the NP decay spectrum where the value in each given energy and the angular bin is now given by the double differential distributions provided in \fulleqref{4-body-decay}. 

\item We also use the binned data from the Monte-Carlo simulations employed by the TWIST collaboration and the theoretical distribution of the three-body muon decay (SM) to obtain an estimate of a bin-by-bin shape function that determines the effects of radiative corrections and detector effects. This shape function allows us to correct for the effects not included in our theoretical distributions. 

\item Finally, we find the preferred value of the NP parameter $g_{\phi l}$ by optimizing the measure $\chi^2$ defined as
     \begin{equation}
        \chi^2(g_{\phi l},\alpha) = N_{obs}^{fid} \sum_{x,y}^{n} \Bigg( \frac{F(g_{\phi l},\alpha;x,y)-\mathcal{G}(x,y)}{\mathcal{G}(x,y)} \Bigg)^2
        \label{chisq}
   \end{equation}
   where $N_{obs}^{fid} \sim 45 \times 10^6$ is the total number of events observed by the TWIST collaboration inside the fiducial region. The functions $F$ and ${\cal G}$ used to define the $\chi^2$ measure are as follows: assuming a light BSM as described by our model, $F$ contains the SM and NP contributions scaled by a weight factor $\alpha$. This is done as we are dealing with probability distributions instead of absolute numbers and the mentioned parametrization implies that the probability distribution $F$ is unit normalized.  Formally, we define the function $F$ as, 
   \begin{equation}
       F(g_{\phi l},\alpha;x,y) = (1-\alpha)f_{3b}(x,y) + \alpha f_{4b}(g_{\phi l},\alpha;x,y)\, ,
   \end{equation}
with $f_{3b}(x,y)$, $f_{4b}(x,y)$, being the normalized binned data for the energy and angle of the outgoing $e^-(e^+)$ generated via the differential decay distribution of Eqs. \ref{3-body-decay} and \ref{4-body-decay} respectively. We immediately note that the optimization parameter, $\alpha$, appearing in the $\chi^2$ measure and the probability distribution $F$ cannot be negative, by definition. 
Finally, the data to which we fit in order to optimize the strength of NP interaction is contained in ${\cal G}$ which is defined as
\begin{equation}
   \mathcal{G}(x,y) = \frac{f_{3b}(x,y) . d_{T}(x,y)}{f_{T}(x,y)}\, , 
\end{equation}
with $d_T(x,y)$ and $f_T(x,y)$ being the data and fit values with the same binning along x and y direction as obtained from the experimental side. 

\item With this construction, we ensure that all the radiative corrections and detector effects are properly cancelled inside the function $\mathcal{G}(x,y)$, as the function $F(g_{\phi l},x,y)$ contains only the leading order effects following Eqs. \ref{3-body-decay} and \ref{4-body-decay}.

\item As a last step, we relate the strength of the NP operator $g_{\phi l}$ to the parameter $\alpha$ appearing in the $\chi^2$ measure as follows
\begin{equation}
\alpha 
\approx  \frac{n^{\rm fid}_{4b}}{n^{{\rm fid}}_{3b}}\cdot \frac{g_{\phi l}^2}{m_l^2 \pi^2} \cdot \frac{ \mathcal{F}(m_{\mu},m_e,m_{\phi})}{m_\mu^5 }
\label{alpha-gphil-conversion}
\end{equation}
with $n^{\rm fid}_{3b,4b}$ being the fraction of events (normalized) surviving the fiducial cuts for both SM and NP distributions.
\end{itemize}

\begin{figure}[t]
	\centering
		\includegraphics[width=0.47\textwidth]{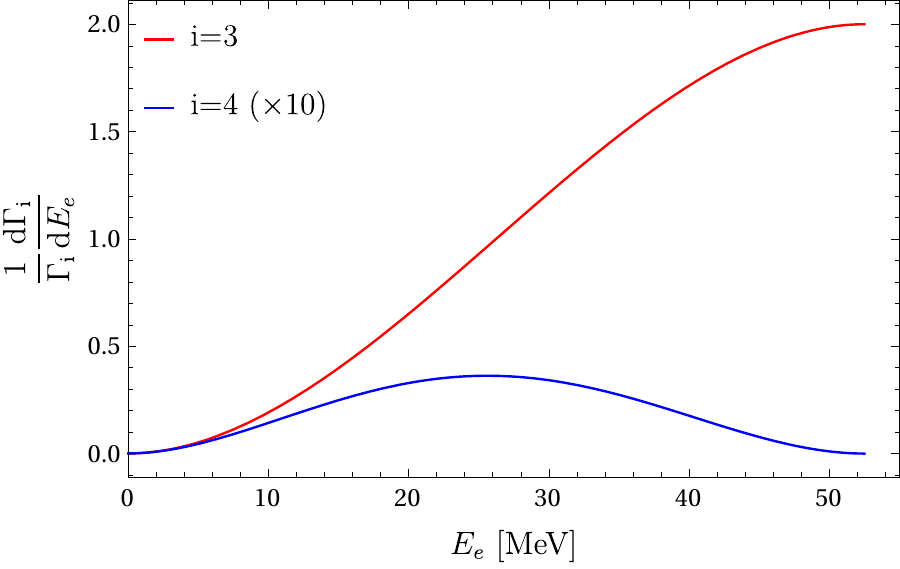}
	\includegraphics[width=0.47\textwidth]{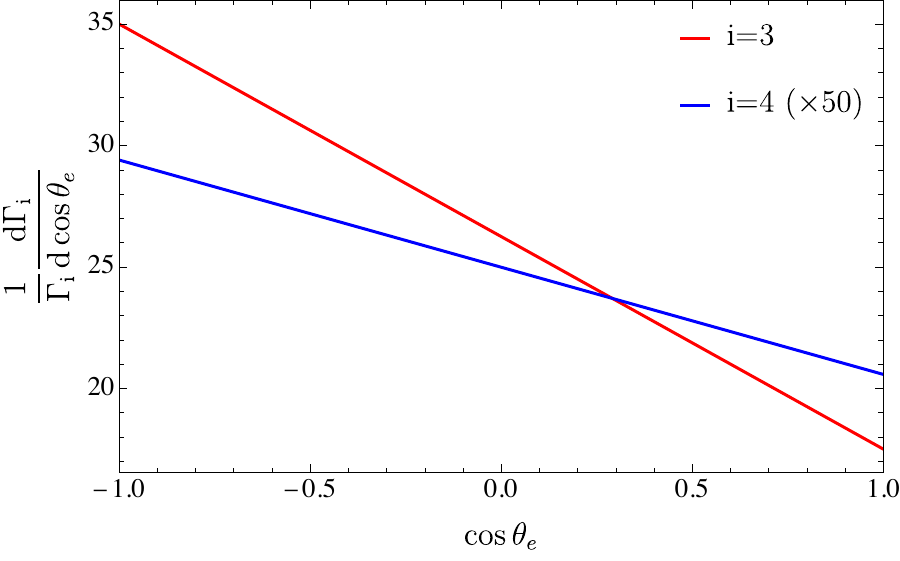}
		\caption{Marginal probability distribution for the energy (left) and angle (right)  normalized to the total decay width for muon decay in SM (in red)
and in BSM only (in blue), respectively. These distributions are obtained by marginalizing the double differential distributions in Eqs. \ref{3-body-decay} and \ref{4-body-decay} over the other variable. For the BSM distributions, we have assumed $m_{\phi} = m_e/100$. }
		\label{fig:marginal_decay}
\end{figure}

Before describing the results of our analysis for the double differential spectrum, it would be instructive to also look at the electron's energy ($E_e$) and angular distribution ($\cos\theta_e$) independently, by marginalizing over the other variable, as shown in \fullfigref{fig:marginal_decay}. Both these marginal distributions are normalized by their corresponding total decay widths in the SM and the NP case. Note that we have multiplied the NP distributions by a factor of $\sim {\cal O}(10^2)$ for the sake of presentation, as the four-body decay spectrum is suppressed by phase space as well as the scale $m_l$. For the energy distribution of the emitted electron (or positron), a distinct shift in the peak can be observed for the spectrum obtained when an additional scalar $\phi$ is emitted along with the lepton. In the case of SM, the energy of the positron emitted always peaks at $m_{\mu}/2$, while in the presence of an additional $\phi$, the maximum electron(or positron) energy is smeared and comes out to be at most $m_{\mu}/4$. This is because when an
additional scalar $\phi$ particle is emitted, $m_{\mu}/2$ amount of energy is being equally
shared between the $e^{\pm}$ and $\phi$ (see \fulleqref{eq:integral-limits}), as opposed to the SM case, where $e^{\pm}$ itself carries the total $m_{\mu}/2$ energy, parting the rest among the two neutrinos.
Along the $\cos\theta_e$ axis, we see the SM background spectrum falls off faster than the four-body signal spectrum as $\cos\theta_e$ goes from
-1 to 1. However, in both cases, the positron is emitted preferentially along the muon spin direction i.e around, $\cos\theta_{e} = -1$.\footnote{Note that, in experiments, it is $\mu^+$ which is used for studying the decay spectrum, as $\mu^-$ becomes quickly bounded in the atoms of the target material, thus the direction of polarization would be along the negative z-axis.} 

  \begin{figure}[!htb]
	
		\includegraphics[width=0.47\textwidth]{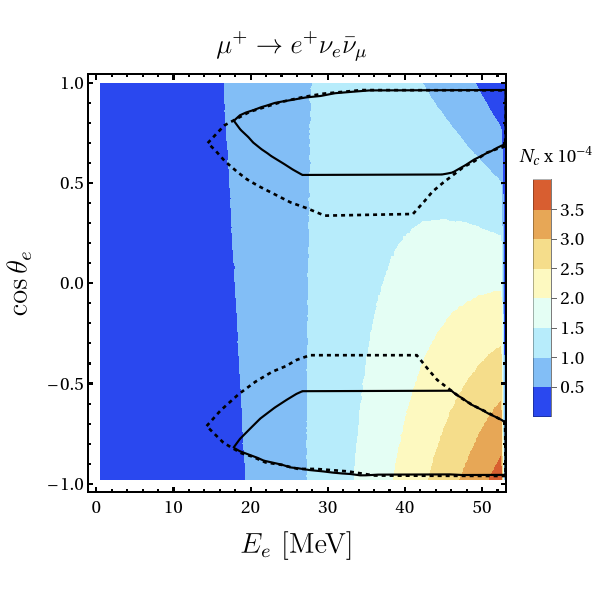}
			\includegraphics[width=0.47\textwidth]{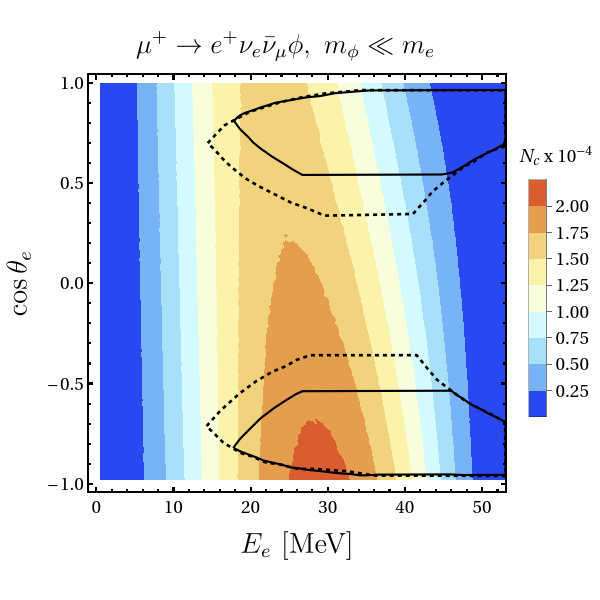}
		\caption{The spectra of polarised muon decays in the case of SM (left) and with an additional scalar (right) in the energy-angle plane. The color legends carry the number of event counts normalized to the total number of events generated following the probability distribution function of \fulleqref{3-body-decay} and \ref{4-body-decay} respectively. The probability distribution functions are normalized to the total number of events generated. The solid black contours in the upper and lower regions denote the fiducial region used in the TWIST experiment and the dashed ones are the proposed area in this analysis.}
		\label{fig:differential_decay}
\end{figure}
The spectra obtained for the polarized muon decay in the energy-angle plane for the SM and the four-body decay mode are shown in \fullfigref{fig:differential_decay}. Here, in the color axis, we represent the number of events in each bin normalized to the total number of events generated.\footnote{We have generated $854 \times 10^6$ events for both the cases keeping in line with the data for set number 87 used in the TWIST experiment \cite{TWIST:2011egd, TWIST:2011aa}. The data from pseudo-experiments is generated following the probability distributions for the three-body and four-body spectra and imposing the same fiducial cuts as specified by the TWIST Collaboration. Furthermore, all the datasets are normalized such that the sum of counts in the fiducial regions is 1.}
The regions within the black contours in \fullfigref{fig:differential_decay} correspond to the fiducial regions used in the experiment. From our knowledge of the marginal distributions, we see a clear shift in the distribution for the four-body decay. The probability distributions are normalized to the total number of events generated in each case, however, one must keep in mind that the four-body distribution will be phase-space suppressed yielding a suppression by around $10^{-4}$ in comparison to the SM case. Notably this differential distribution depends on the mass of the scalar that we employ as discussed previously in Sec.~\ref{sec:four-body}. Here we have used $m_{\phi} = m_e/100$ for the distribution shown in \fullfigref{fig:differential_decay}. 

\begin{figure}
         \centering
          \includegraphics[width=0.47\textwidth]{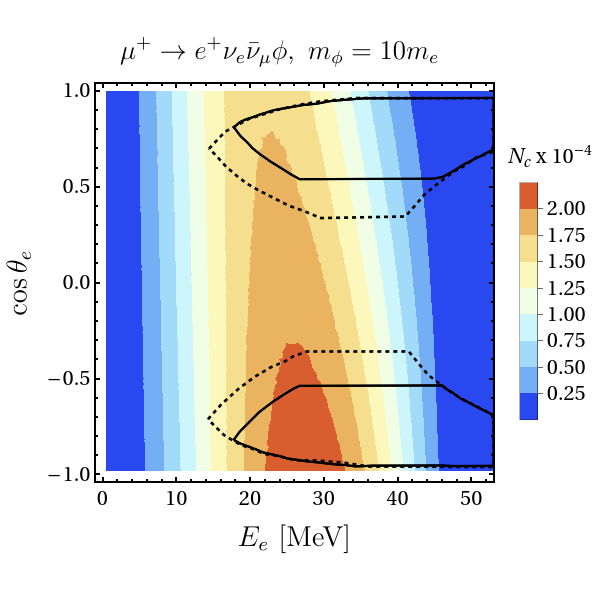}
           \includegraphics[width=0.47\textwidth]{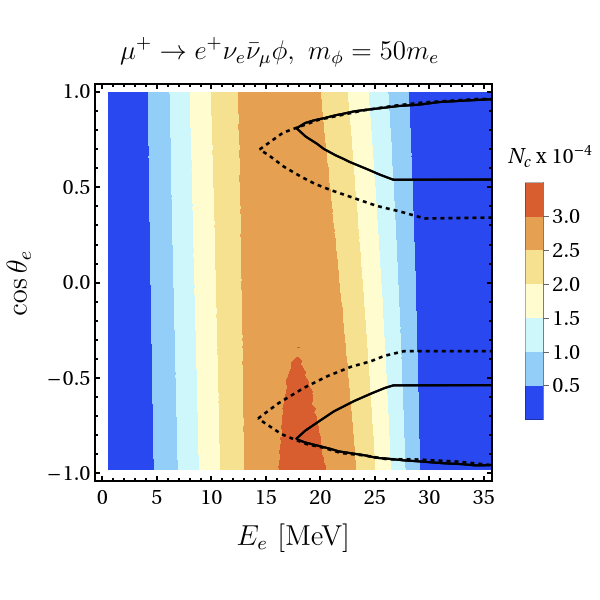}
         \caption{The electron recoil energy distribution with the variation of ALP mass,  $ m_{\phi} = 10\, m_e$ (left) and $50\, m_e$ (right). The position of the maxima shifts towards the left with increasing ALP mass. The probability distribution functions are normalized to the total number of events generated. The black contours in the upper and lower regions denote the fiducial region used in the TWIST experiment. The black dashed lines show the proposed region for the analysis.}
         \label{fig:ALP-mass-vary}
     \end{figure}

The spectra obtained by varying the mass of the NP scalar are also shown in \fullfigref{fig:ALP-mass-vary}, in particular for two specific values $m_\phi =10 m_e$ and $m_\phi = 50 m_e$. From \fullfigref{fig:ALP-mass-vary}, we can clearly see that as the NP particle is taken to be heavier, the allowed energy available to the outgoing electron (positron) is reduced, inclined to our expectation.\footnote{It is important to note that for the case where the ALP is significantly heavier than the electron, its decay $\phi \to e^{+}\, e^{-}$ is kinematically allowed. However, as we assume the dominant coupling of the ALP to the dark sector, this decay probability is suppressed in our setup. Nevertheless, even if the constraint is relaxed for a more realistic analysis, we would like to point out that with the only NP operator that we use, this probability is still suppressed due to phase space.} We also note that with the same set of fiducial cuts, a lower number of events will survive in the allowed region for the case of the heavier ALP.

Knowing the binned data from our simulations both for the three-body and four-body decays, i.e. $f_{3b}$ and $f_{4b}$, along with the binned data from the experimental simulations and the data, $f_{T}$ and $d_T$, we minimize the measure defined in \fulleqref{chisq} to obtain the best-fit value for the parameter $\alpha$ that describes the strength of the NP that can be accommodated.

\section{Constraints on the new physics coupling}
\label{sec:constraints}

We now determine the constraints on the NP coupling utilizing the data provided by the TWIST collaboration. By minimizing the $\chi^2$ measure, we aim to obtain bounds on our NP coupling at 68\% confidence level (CL), for various values of the ALP mass.
For an ultra-light ALP with mass $m_{\phi} = m_e/100$, we find that the best-fit value of the fit parameter is compatible with the SM expectation ($\alpha = 0$), thereby allowing for only a minimal NP effect at a 68\% CL, giving $g_{\phi l} \leq 0.0032$. To get to this bound, we have used the total number of events that are contained in the fiducial region for the three-body and four-body decays respectively, as 45 $\times 10^{6}$ for the SM case while we take the total number of events surviving the fiducial cuts from our Monte-Carlo simulations for the four-body case.

\begin{table}[!htb]
\centering
\begin{tabular}{|c|c|c|c|}
\hline
     ALP Mass [MeV] &  Events in the fiducial area ($N_{4b}^{fid}$)  & $\alpha$ at $68\%$ CL & $g_{\phi l}$  \\
     \hline
      $m_e/100$ &  $238 \times 10^6$ & $1.50 \times 10^{-4}$ & $0.00316$\\
     $m_e$ &  $238 \times 10^6$ & $1.50 \times 10^{-4}$ & $0.00316$ \\
     $10 m_e$ & $226 \times 10^6$& $1.24 \times 10^{-4}$ & $0.00382$ \\
     $50 m_e$ & $94 \times 10^6$ & $3.0 \times 10^{-5}$ & $0.00604$ \\
     \hline
\end{tabular}
 \caption{The NP coupling obtained from different ALP masses from the TWIST differential analysis using \fulleqref{chisq}. }
 \label{lambdaeff_with_mass_table}
\end{table} 

Table~\ref{lambdaeff_with_mass_table} displays how the strength of the NP interaction modifies with the mass of the ALP. 
We find that the constraint gets weaker with increasing ALP masses. This can be understood as the heavier the mass of the ALP, the smaller the probability of it decaying into such a four-body final state, see Appendix~\ref{appendix-A} for more details. Moreover, the energy distribution of the emitted electron shifts to much lower values, implying the number of events surviving inside the experimentally accessible region reduces, further guaranteeing the observed effect. 

This leads us to an interesting consideration: modify the fiducial cuts in our simulations with the goal of enhancing the NP contribution. In principle, this would allow for better constraints on the strength of the proposed NP interaction. We do this by keeping in mind that not only we would like to obtain an observational region where the contributions from NP are significant and comparable to those from the SM,  additionally, we would like to ensure that this region is realistically achievable in experiments.
 Noting that the present experimental efforts are restricted by reconstructing the small transverse momenta of the particles, we choose the cuts as follows: $ 10.0 ~ {\rm MeV} < p_t < 39.0$ MeV, $p_l > 10.0$ MeV and $0.34 < |\cos\theta| < 0.96$. The dashed region in Figure~\ref{fig:ALP-mass-vary} represents the proposed set of cuts. The fraction of events that survive these proposed cuts are tabulated in Table~\ref{Table:changing_cuts}. Notice that even with the minimal modification to the TWIST cuts, the population of NP events in the proposed regions increases roughly by a factor of 2 while the SM is not affected so strongly gaining only by about $20\%$. We hope that with the proposed set of cuts, the NP operator under study can be more strongly constrained through a bin-by-bin analysis as shown here. 

\begin{table}[!htb]
\centering
\begin{tabular}{|C{3cm}|C{2cm}|C{2cm}|C{2cm}|C{2cm}|}
\hline
      &  \multicolumn{2}{c|}{TWIST Cuts}  & \multicolumn{2}{c|}{Proposed Cuts}  \\
     \cline{2-5}
     & & & & \\
     ALP Mass [MeV]  & $N_{3b}^{fid}/N_{tot}$ & $N_{4b}^{fid}/N_{tot}$ & $N_{3b}^{fid}/N_{tot}$ & $N_{4b}^{fid}/N_{tot}$ \\
     & & & & \\
     \hline
     $m_e/100$  & -  & $0.279$ & - & $0.432$\\
     $10 m_e$   &  $0.342$  & $0.265$ &$0.439$ & $0.420$ \\
     $50 m_e$   &  -  & $0.110$ & -   & $0.223$  \\
     \hline
\end{tabular}
 \caption{The fraction of events that survived the fiducial cuts used in the {\rm TWIST} analysis and those that survived the proposed cut area. $N_{tot} = 854 \times 10^6$ in both the cases. The numbers in each column represent the fractional count inside the solid {\rm (TWIST)} and dashed {\rm (Proposed)} contours, see \fullfigref{fig:differential_decay} and \fullfigref{fig:ALP-mass-vary}, in the three-body and four-body distributions.}
 \label{Table:changing_cuts}
\end{table}

Next, we also consider existing constraints on ALPs originating from leptonic decays of various mesons, see Refs.~\cite{Charm:1985, Bjorkeroth:2018dzu, PIONEER:2022yag, Poblaguev:2002ug, Altmannshofer:2022ckw, Calibbi:2020jvd} and references therein.
\footnote{
In this context, lepton flavor-violating (LFV) coupling 
among the electron, muon, and axion, $\mu \rightarrow e \phi$ 
has also been considered. This process differs 
significantly from others in terms of the effective 
operator and phase space of the decay. Utilizing a high-
purity germanium detector, Ref.\cite{Bilger:1998rp} 
examined muon decay within the mass range of $103-105$ 
MeV, establishing a limit of $5.7 \times 10^{-4}$. 
The TWIST experiment conducted a broader search for 
masses up to $\simeq 80$ MeV, 
accommodating non-zero anisotropies, resulting in an 
upper limit of $2.1 \times 10^{-5}$ for massless axions
\cite{TWIST:2014ymv}. For a comprehensive overview of 
its current state, please refer to 
\cite{Calibbi:2020jvd}.
}
Here it is relevant to point out that these may not be a one-to-one comparison to the particular operator discussed here for two reasons. First, in meson decays like $\pi^{\pm} \to e^{\pm} \nu_e \phi$, it is not possible to disentangle ALP-lepton interaction vertex from the ALP-quark interactions. Second, the dominant search channel in these set-ups utilizes ALP decays to visible states, which is different from the operator we intend to study.\footnote{To the best of our knowledge, the only scenario where axions contributing to invisibles is considered was conducted by the PIENU collaboration for the study of charged pion decays~\cite{PIENU:2021clt, PIENU:2020loi}.} A brief discussion about various electroweak completion scenarios that can lead to the effective operator we consider is presented in Appendix~\ref{appendix-B}. From this, it is clear that while these decays might be relevant in understanding which part of the phase space is constrained through the analysis of TWIST data considered here, these should not necessarily be treated on the same footing. In fact, we find that the region constrained by the TWIST experiment exhibits a lag compared to other bounds derived from meson decays and electron beam dump experiments.       

Finally, some comments about the ALP stability such that it remains undetected within the extent of the collider are in order.
With the only NP interaction as given in \fulleqref{new_phi_operator}, the ALP decays primarily to a four-body final state containing two electrons and two neutrinos. We find that 
such a decay width of the ALP to SM fermions is of the order of  $10^{-26}$ GeV.  Additionally, decay to a pair of electrons is also viable but would then be further loop suppressed. This smallness of the decay width suggests that, within the current set-up, the ALP could very well be considered to be stable within the length of the collider.
From a model-building point of view, this could be achieved by introducing a coupling between the ALP and a dark sector, assuming that this coupling is stronger than the ALP's interactions with the visible sector, as governed by the $SU(2)_L$ gauge coupling~\cite{Nelson:2011sf, Arias:2012az}. Although this might be interesting in its own right, exploring the dedicated phenomenology of the dark sector is beyond the scope of the current study.

\section{Conclusions}
\label{sec:conclusions}
In summary, 
we focus on ALPs produced in association with electrons and neutrinos within the muon decay process. By employing a 
$d=7$ effective operator,
we analyze 
the possible decay width of muons into a four-body channel. Drawing upon the dataset furnished by the TRIUMF Weak Interaction Symmetry Test (TWIST) experiment, we derive constraints on ALP-lepton couplings spanning a spectrum of masses, extending up to $m_{\mu}/4$ 
as constrained by kinematics. 
Assuming the ALP decays predominantly to invisible final states, we find constraints on the effective scale of the new physics operator. Although we find weaker bounds than the ones coming from charged meson decays, the set-up we consider cleanly constrains only the ALP-lepton coupling in contrast to the couplings constrained by meson decay processes, thus providing a complementary search strategy. 
As a byproduct, utilizing our simulations, we find that even by extending the fiducial cuts within the TWIST set-up very minimally, one can hope to enhance the experimental sensitivity towards such a NP, if it exists. 

Keeping in mind the forthcoming experimental endeavors involving muon beams, 
we would like to obtain an order of estimate from the existing dataset necessary to detect such a decay at future experimental facilities, for eg., MEG-II~\cite{MEGII:2021fah, Meucci:2022qbh}.
This entails determining the number of events required for observing such a decay in a statistically significant manner. 
From the partial decay width, we expect that reaching at least a count of $\mathcal{O}(10^{12})$ muons in a beam would be necessary to have a non-zero probability of the muon to decay to such a four-body final state. 
We hope that to be able to observe or rule out such a proposed four-body final state would 
allow us to constrain the parameter space for light ALPs that can potentially remain invisible.

\section*{Acknowledgements}
The authors sincerely acknowledge Prof. Arthur Olin and others from the TWIST collaboration at TRIUMF for kindly providing us with the data sets and fitting results. We thank Tuhin S. Roy for helpful discussions throughout the extent of the project and Triparno Bandyopadhyay for his useful comments on the manuscript. The work of SM is supported by the IOE-IISc fellowship. AB and SM acknowledge the hospitality provided by TIFR where a substantial part of this work was completed.

\appendix
\section{Functional dependence of four-body decay width}
\label{appendix-A}
The full functional form of  $\mathcal{F}(m_{\mu},m_e, m_{\phi})$ used in \fulleqref{eq:Gamma4_function} and taken in the analysis are given as,
\begin{align}
   \mathcal{F} & = \frac{1}{6720}\Bigg[840 \, m_{\phi}^2 \left(\log (2 m_{\phi})-\log \left(\frac{m_{\phi }^2}{m_{\mu }}+2 m_{\phi }+ f(m_{\mu},m_{\phi})\right)\right) {\cal I}_{1}(m_{\mu},m_e,m_{\phi}) +\frac{840\, m_{\phi}^2}{m_{\mu }^3} \nonumber \\
   &\quad \log \left(m_{\phi }^2+2 m_{\mu } m_{\phi }+ m_{\mu } f(m_{\mu},m_{\phi})\right){\cal I}_{2}(m_{\mu},m_{e},m_{\phi}) +\frac{m_{\phi}^2}{m_{\mu }^6}  g(m_{\mu},m_{\phi}) {\cal I}_3(m_{\mu},m_{e},m_{\phi}) \nonumber\\
&\quad
-\frac{840\,m_{\phi }^2}{m_{\mu}^3}\, \log(2 m_{\mu}^2) {\cal I}_2(m_{\mu},m_{e},m_{\phi}) +\frac{g(m_{\mu},m_{\phi})}{m_{\mu }^6} {\cal I}_4(m_{\mu},m_e,m_{\phi}) \Bigg]\, ,
\end{align}
with $f(m_{\mu},m_{\phi})= \sqrt{\frac{m_{\phi }^3 \left(4 m_{\mu }+m_{\phi }\right)}{m_{\mu }^2}}$ and $g(m_{\mu},m_{\phi})=\frac{\left(m_{\mu }^2-m_{\phi }^2\right)}{m_{\mu }}$.
The functions ${\cal I}_{1,2,3,4}$ appearing in the above equation are polynomials in the masses of the particles involved and are given as:
\begin{align}
{\cal I}_{1}&= (\left(m_{\mu }-2 m_{\phi }\right) \left(m_{\mu }+m_{\phi }\right){}^2-2 m_e^2 \left(m_{\mu }+2 m_{\phi }\right)) \left(m_{\mu }-2 m_{\phi }\right)^2  \\
{\cal I}_2 &= m_{\mu }^8+7 m_{\phi }^2 m_{\mu }^6+5 m_{\phi }^4 m_{\mu }^4-5 m_{\phi }^6 m_{\mu }^2-2 m_{\phi }^8 -2 m_e^2 m_{\mu}\, g(m_{\mu},m_{\phi}) \left(m_{\mu }^4+3 m_{\phi }^2 m_{\mu }^2+m_{\phi }^4\right)  \\
{\cal I}_3 &=-1680\, m_{\mu }^{10}-2240\, m_{\phi } m_{\mu }^9+4340\, m_{\phi }^2 m_{\mu }^8-3640\, m_{\phi }^3 m_{\mu }^7-3122\, m_{\phi }^4 m_{\mu }^6 +504\, m_{\phi }^5 m_{\mu }^5 \nonumber \\
&\quad +2639\, m_{\phi }^6 m_{\mu }^4-402\, m_{\phi }^7 m_{\mu }^3+73\, m_{\phi }^8 m_{\mu }^2+70\, m_e^2 (24\, m_{\mu }^6+68 m_{\phi } m_{\mu }^5-104\, m_{\phi }^2 m_{\mu }^4 -6\, m_{\phi }^3 m_{\mu }^3 \nonumber \\
&\quad -17\, m_{\phi }^4 m_{\mu }^2+2\, m_{\phi }^5 m_{\mu }-m_{\phi }^6 ) m_{\mu }^2+6 m_{\phi }^9 m_{\mu }-3\, m_{\phi }^{10} \\
{\cal I}_{4} &= 17\, m_{\mu }^{12}+1326\, m_{\phi }^2 m_{\mu }^{10}+3979\, m_{\phi }^4 m_{\mu }^8 -746\, m_{\phi }^6 m_{\mu }^6-1971\, m_{\phi }^8 m_{\mu }^4-88\, m_{\phi }^{10} m_{\mu }^2+3\, m_{\phi }^{12}  \nonumber \\
& \quad -70\, m_{\mu }^2 m_e^2 \left(m_{\mu }^8+28 m_{\phi }^2 m_{\mu }^6-28\, m_{\phi }^6 m_{\mu }^2-m_{\phi }^8\right) 
\end{align}

\section{Electroweak considerations for the effective operator}
\label{appendix-B}
Up to this point of the study, we have described a compelling and minimalistic scenario that accounts for the four-body decay of muons. Its simplicity allows it to evade detection in the numerous searches for new physics. In this appendix, we explore the theoretical construction for \fulleqref{new_phi_operator}, although in terms of higher dimensional irrelevant operators. Our focus here is to simply show that there could be numerous such possibilities that can be constructed. 

We first note that the operator studied here is similar to the one earlier explored in the context of $W$ mass anomaly in Ref.~\cite{Bandyopadhyay:2022bgx}, with the light scalar coupled to SM quark fields. Above the regime of Fermi theory, one can start from the gauge-covariant kinetic term in the lepton sector of the SM-Lagrangian and allow a field-dependent transformation of the lepton doublet as:
\begin{equation}\label{eq:redefinition}
	{\ell}_{L/R} \ \rightarrow \ \exp\left(i x^i_{L/R}t_i \phi \right) {\ell}_{L/R} 
\end{equation}
where $t_i$ are $SU(2)$ generators. Note that $t^{1,2}$ or rather $t^{\pm}$ breaks U(1) electromagnetism and hence terms proportional to $t^{1,2}$ are not allowed. Similarly, although $t_3$ generator would break electroweak (EW) invariance, however, we keep it for now.
In the present setup, with only the left-handed currents associated with W-bosons. 
If we start by adding more familiar constructions of ALPs, i.e. $\delta L = \partial_{\mu}\phi(\bar{\nu}_L \gamma^{\mu}\nu_L - \bar{l}_L \gamma^{\mu} l_L )$ then the field redefinition in \fulleqref{eq:redefinition} eliminates this particular operator but gives rise to $\phi (\bar\mu \gamma^{\nu} P_L \nu_{\mu}) W_{\nu}^{-}$ with appropriate coefficients. 
One can easily show that the amplitude for observables remains unaltered under this redefinition.
Considering the explicit ALP interaction term in mind, $\delta{\cal L} \supset \partial_{\mu}\phi \, j^{\mu}$, the most general muon current associated with a light (pseudo)scalar, consists of $\bar{\mu}\gamma^{\nu}\mu$, $\bar{\mu}\gamma^{\nu}\gamma_5\mu$ and $\bar{\nu}_{\mu}\gamma^{\nu} P_L \nu_{\mu}$ terms. Note that, these will not be independent under electroweak invariance. But, in general, each lepton coupling can arise separately in a weak-invariant theory by including their corresponding currents. Some possible examples of such ultraviolet (UV) constructions were also explored in Ref.~\cite{PhysRevLett.60.1793}. After integration by parts of the Lagrangian, we again end up with the term $\phi (\bar\mu \gamma^{\nu} P_L \nu_{\mu}) W_{\nu}^{-}$. However, note that this four-point interaction would vanish when the general muon current respects the electroweak symmetry \cite{PhysRevLett.60.1793, Altmannshofer:2022ckw, Lu:2023ryd}. This interaction is independent of $m_{\mu}$ and may be crucial in constraining the light ALP. 

Furthermore, $\phi$ can be associated with the massive gauge field itself through the $(F_{\mu\nu})^2$ term.
In this case, we can write more familiar $SU(2)_L$ gauge kinetic term coupling with $\phi$ at $d=5$ level,
\begin{equation}
\begin{split}
    \mathcal{L} &\supset -\frac{1}{4g_{2}^{2}}W_{\mu\nu}^{a}W^{\mu\nu a} + \frac{1}{4g_{2}^{2}}\frac{\kappa}{M}\phi W_{\mu\nu}^{a}W^{\mu\nu a} \\
    &=  -\frac{1}{4g_{2}^{2}}\left( 1 - \frac{\kappa}{M}\phi \right)W_{\mu\nu}^{a}W^{\mu\nu a},
\end{split}
\end{equation}
\\
with $M$ being the mass of the messenger field of the intermediate effective theory and $\kappa$ is the dimensionless coupling. The fraction $\frac{\kappa}{M} \simeq \frac{1}{16\pi^2}\frac{g_{2}^{2}}{M_Q}$ in the case if the messenger field is a massive fermion with mass $M_{Q}$ and $\frac{\kappa}{M} \simeq \frac{1}{16\pi^2}\frac{g_{2}^{2}}{M_{\Phi}^{2}}$ if the messenger field is a massive complex scalar field. The inclusion of the $\phi W_{\mu\nu}^{a} W^{\mu\nu a}$ effectively shifts the coupling constant $g_2$ as,
\begin{equation}
    \frac{1}{g_{2}^{2}} \rightarrow \frac{1}{g_{2}^{2}}\left( 1 - \frac{\kappa}{M}\phi \right).
\end{equation}
We can once again rescale the massive gauge terms: $W_{\mu}^a \rightarrow g_2 W_{\mu}^a$. Using equations of motion, we can express the gauge field $W_{\mu}^a$ up to order $\mathcal{O}(1/m_W^2)$ as, $W_{\mu}^a \simeq \frac{g_2^{\prime}}{m_W^2}(\bar{\psi}\gamma_{\mu}t^a\psi)$. Subsequently, the effective operator after integrating out the massive gauge boson terms becomes,
\begin{equation}
    \mathcal{L} \supset  \frac{g_{2}^{2}}{2 m_W^{2}}\left( 1 + \frac{\kappa}{M}\phi \right)(\bar{\psi}\gamma_{\mu}t^a\psi)(\bar{\psi}\gamma^{\mu}t^a\psi).
\end{equation}
The low energy effective muon decay operator in the Standard Model has a coefficient $G_F \sim \frac{g_2^2}{m_W^2}$.
The inclusion of the $\phi$ field along with the heavy messenger field gives a correction to this coefficient $G_F$, either as a
correction to $g_2$ or to $m_W$. The correction to the masses $m_W$ of the $SU(2)$ gauge fields could occur in models where the
messenger field is an SM complex scalar acquiring vacuum expectation value (\textit{vev}). 
Another possible UV scenario which results in this kind of EFT is to design a scalar
potential (using only marginal and relevant operators) with the $\phi$ field and the doublet Higgs field ($H$) as,
\begin{equation}
V(H,\phi) \ = \ -\mu^2 (H^{\dagger}H) + \lambda  (H^{\dagger}H)^2 - y \phi H^{\dagger}H
\end{equation}
Note that, this potential is just a simplest example and falls under the multitude of hidden sector Higgs-portal scenarios \cite{Patt:2006fw, Arcadi:2019lka, Arcadi:2021mag}.
The minimum of the potential is found at a finite value, which is shifted from the original vacuum expectation value and linearly depends on $\phi$. We find that the replacement of the EW \textit{vev} (and
therefore of $G_F$) by its $\phi-$dependent value allows us to re-derive the low energy EFT \cite{Bandyopadhyay:2021wbb}.
Thus we see how the effective four-fermion-scalar operator could possibly be generated from higher dimensional irrelevant operators and can be used for the low-energy effective study that we perform. Further considerations regarding EW T-parameter and/or flavour considerations rely on the specifics of a concrete UV model and are beyond the scope of this study.

\bibliography{reference}

\providecommand{\href}[2]{#2}\begingroup\raggedright\begin{thebibliography}{10}

\bibitem{Weinberg:1977ma}
S.~Weinberg, \emph{{A New Light Boson?}},
  \href{https://doi.org/10.1103/PhysRevLett.40.223}{\emph{Phys. Rev. Lett.}
  {\bfseries 40} (1978) 223}.

\bibitem{Wilczek:1977pj}
F.~Wilczek, \emph{{Problem of Strong $P$ and $T$ Invariance in the Presence of
  Instantons}}, \href{https://doi.org/10.1103/PhysRevLett.40.279}{\emph{Phys.
  Rev. Lett.} {\bfseries 40} (1978) 279}.

\bibitem{Kim:2008hd}
J.E.~Kim and G.~Carosi, \emph{{Axions and the Strong CP Problem}},
  \href{https://doi.org/10.1103/RevModPhys.82.557}{\emph{Rev. Mod. Phys.}
  {\bfseries 82} (2010) 557} [\href{https://arxiv.org/abs/0807.3125}{{\ttfamily
  0807.3125}}].

\bibitem{Marsh:2015xka}
D.J.E.~Marsh, \emph{{Axion Cosmology}},
  \href{https://doi.org/10.1016/j.physrep.2016.06.005}{\emph{Phys. Rept.}
  {\bfseries 643} (2016) 1} [\href{https://arxiv.org/abs/1510.07633}{{\ttfamily
  1510.07633}}].

\bibitem{Brivio:2017ije}
I.~Brivio, M.B.~Gavela, L.~Merlo, K.~Mimasu, J.M.~No, R.~del Rey et~al.,
  \emph{{ALPs Effective Field Theory and Collider Signatures}},
  \href{https://doi.org/10.1140/epjc/s10052-017-5111-3}{\emph{Eur. Phys. J. C}
  {\bfseries 77} (2017) 572}
  [\href{https://arxiv.org/abs/1701.05379}{{\ttfamily 1701.05379}}].

\bibitem{Jaeckel:2010ni}
J.~Jaeckel and A.~Ringwald, \emph{{The Low-Energy Frontier of Particle
  Physics}},
  \href{https://doi.org/10.1146/annurev.nucl.012809.104433}{\emph{Ann. Rev.
  Nucl. Part. Sci.} {\bfseries 60} (2010) 405}
  [\href{https://arxiv.org/abs/1002.0329}{{\ttfamily 1002.0329}}].

\bibitem{Ringwald:2012hr}
A.~Ringwald, \emph{{Exploring the Role of Axions and Other WISPs in the Dark
  Universe}}, \href{https://doi.org/10.1016/j.dark.2012.10.008}{\emph{Phys.
  Dark Univ.} {\bfseries 1} (2012) 116}
  [\href{https://arxiv.org/abs/1210.5081}{{\ttfamily 1210.5081}}].

\bibitem{Essig:2013lka}
R.~Essig et~al., \emph{{Working Group Report: New Light Weakly Coupled
  Particles}},  in \emph{{Snowmass 2013}: {Snowmass on the Mississippi}}, 10,
  2013 [\href{https://arxiv.org/abs/1311.0029}{{\ttfamily 1311.0029}}].

\bibitem{Peccei:1977hh}
R.D.~Peccei and H.R.~Quinn, \emph{{CP Conservation in the Presence of
  Instantons}}, \href{https://doi.org/10.1103/PhysRevLett.38.1440}{\emph{Phys.
  Rev. Lett.} {\bfseries 38} (1977) 1440}.

\bibitem{Peccei:1977ur}
R.~Peccei and H.~Quinn, \emph{{Constraints Imposed by CP Conservation in the
  Presence of Instantons}},
  \href{https://doi.org/10.1103/PhysRevD.16.1791}{\emph{Phys. Rev. D 16 (1977)
  1791} (1977) }.

\bibitem{Konaka:1986cb}
A.~Konaka et~al., \emph{{Search for Neutral Particles in Electron Beam Dump
  Experiment}}, \href{https://doi.org/10.1103/PhysRevLett.57.659}{\emph{Phys.
  Rev. Lett.} {\bfseries 57} (1986) 659}.

\bibitem{Riordan:1987aw}
E.M.~Riordan et~al., \emph{{A Search for Short Lived Axions in an Electron Beam
  Dump Experiment}},
  \href{https://doi.org/10.1103/PhysRevLett.59.755}{\emph{Phys. Rev. Lett.}
  {\bfseries 59} (1987) 755}.

\bibitem{Bjorken:1988as}
J.D.~Bjorken, S.~Ecklund, W.R.~Nelson, A.~Abashian, C.~Church, B.~Lu et~al.,
  \emph{{Search for Neutral Metastable Penetrating Particles Produced in the
  SLAC Beam Dump}}, \href{https://doi.org/10.1103/PhysRevD.38.3375}{\emph{Phys.
  Rev. D} {\bfseries 38} (1988) 3375}.

\bibitem{Bross:1989mp}
A.~Bross, M.~Crisler, S.H.~Pordes, J.~Volk, S.~Errede and J.~Wrbanek, \emph{{A
  Search for Shortlived Particles Produced in an Electron Beam Dump}},
  \href{https://doi.org/10.1103/PhysRevLett.67.2942}{\emph{Phys. Rev. Lett.}
  {\bfseries 67} (1991) 2942}.

\bibitem{Scherdin:1991xy}
A.~Scherdin, J.~Reinhardt, W.~Greiner and B.~Muller, \emph{{Low-energy e+ e-
  scattering}}, \href{https://doi.org/10.1088/0034-4885/54/1/001}{\emph{Rept.
  Prog. Phys.} {\bfseries 54} (1991) 1}.

\bibitem{Tsai:1989vw}
Y.-S.~Tsai, \emph{{PRODUCTION OF NEUTRAL BOSONS BY AN ELECTRON BEAM}},
  \href{https://doi.org/10.1103/PhysRevD.40.760}{\emph{Phys. Rev. D} {\bfseries
  40} (1989) 760}.

\bibitem{Bassompierre:1995kz}
G.~Bassompierre et~al., \emph{{Search for light neutral objects photoproduced
  in a crystal strong field and decaying into e+ e- pairs}},
  \href{https://doi.org/10.1016/0370-2693(95)00628-X}{\emph{Phys. Lett. B}
  {\bfseries 355} (1995) 584}.

\bibitem{Izaguirre:2016dfi}
E.~Izaguirre, T.~Lin and B.~Shuve, \emph{{Searching for Axionlike Particles in
  Flavor-Changing Neutral Current Processes}},
  \href{https://doi.org/10.1103/PhysRevLett.118.111802}{\emph{Phys. Rev. Lett.}
  {\bfseries 118} (2017) 111802}
  [\href{https://arxiv.org/abs/1611.09355}{{\ttfamily 1611.09355}}].

\bibitem{Marciano:2016yhf}
W.J.~Marciano, A.~Masiero, P.~Paradisi and M.~Passera, \emph{{Contributions of
  axionlike particles to lepton dipole moments}},
  \href{https://doi.org/10.1103/PhysRevD.94.115033}{\emph{Phys. Rev. D}
  {\bfseries 94} (2016) 115033}
  [\href{https://arxiv.org/abs/1607.01022}{{\ttfamily 1607.01022}}].

\bibitem{Berlin:2018bsc}
A.~Berlin, N.~Blinov, G.~Krnjaic, P.~Schuster and N.~Toro, \emph{{Dark Matter,
  Millicharges, Axion and Scalar Particles, Gauge Bosons, and Other New Physics
  with LDMX}}, \href{https://doi.org/10.1103/PhysRevD.99.075001}{\emph{Phys.
  Rev. D} {\bfseries 99} (2019) 075001}
  [\href{https://arxiv.org/abs/1807.01730}{{\ttfamily 1807.01730}}].

\bibitem{AristizabalSierra:2020rom}
D.~Aristizabal~Sierra, V.~De~Romeri, L.J.~Flores and D.K.~Papoulias,
  \emph{{Axionlike particles searches in reactor experiments}},
  \href{https://doi.org/10.1007/JHEP03(2021)294}{\emph{JHEP} {\bfseries 03}
  (2021) 294} [\href{https://arxiv.org/abs/2010.15712}{{\ttfamily
  2010.15712}}].

\bibitem{Gori:2020xvq}
S.~Gori, G.~Perez and K.~Tobioka, \emph{{KOTO vs. NA62 Dark Scalar Searches}},
  \href{https://doi.org/10.1007/JHEP08(2020)110}{\emph{JHEP} {\bfseries 08}
  (2020) 110} [\href{https://arxiv.org/abs/2005.05170}{{\ttfamily
  2005.05170}}].

\bibitem{Bauer:2020jbp}
M.~Bauer, M.~Neubert, S.~Renner, M.~Schnubel and A.~Thamm, \emph{{The
  Low-Energy Effective Theory of Axions and ALPs}},
  \href{https://doi.org/10.1007/JHEP04(2021)063}{\emph{JHEP} {\bfseries 04}
  (2021) 063} [\href{https://arxiv.org/abs/2012.12272}{{\ttfamily
  2012.12272}}].

\bibitem{Bauer:2021mvw}
M.~Bauer, M.~Neubert, S.~Renner, M.~Schnubel and A.~Thamm, \emph{{Flavor probes
  of axion-like particles}},
  \href{https://doi.org/10.1007/JHEP09(2022)056}{\emph{JHEP} {\bfseries 09}
  (2022) 056} [\href{https://arxiv.org/abs/2110.10698}{{\ttfamily
  2110.10698}}].

\bibitem{Altmannshofer:2022ckw}
W.~Altmannshofer, J.A.~Dror and S.~Gori, \emph{{New Opportunities for Detecting
  Axion-Lepton Interactions}},
  \href{https://doi.org/10.1103/PhysRevLett.130.241801}{\emph{Phys. Rev. Lett.}
  {\bfseries 130} (2023) 241801}
  [\href{https://arxiv.org/abs/2209.00665}{{\ttfamily 2209.00665}}].

\bibitem{Lu:2022zbe}
C.-T.~Lu, \emph{{Lighting electroweak-violating ALP-lepton interactions at e+e-
  and ep colliders}},
  \href{https://doi.org/10.1103/PhysRevD.108.115029}{\emph{Phys. Rev. D}
  {\bfseries 108} (2023) 115029}
  [\href{https://arxiv.org/abs/2210.15648}{{\ttfamily 2210.15648}}].

\bibitem{Lu:2023ryd}
C.-T.~Lu, X.~Luo and X.~Wei, \emph{{Exploring muonphilic ALPs at muon
  colliders}}, \href{https://doi.org/10.1088/1674-1137/ace424}{\emph{Chin.
  Phys. C} {\bfseries 47} (2023) 103102}
  [\href{https://arxiv.org/abs/2303.03110}{{\ttfamily 2303.03110}}].

\bibitem{TWISTprd2011}
J.F.~Bueno, \emph{{TWIST, Precise measurement of parity violation in polarized
  muon decay}}, {\emph{Phys. Rev. D} {\bfseries 84, 032005} (2011) }
  [\href{https://arxiv.org/abs/1104.3632 [hep-ex]}{{\ttfamily 1104.3632
  [hep-ex]}}].

\bibitem{Nomura:2008ru}
Y.~Nomura and J.~Thaler, \emph{{Dark Matter through the Axion Portal}},
  \href{https://doi.org/10.1103/PhysRevD.79.075008}{\emph{Phys. Rev. D}
  {\bfseries 79} (2009) 075008}
  [\href{https://arxiv.org/abs/0810.5397}{{\ttfamily 0810.5397}}].

\bibitem{Bharucha:2022lty}
A.~Bharucha, F.~Br\"ummer, N.~Desai and S.~Mutzel, \emph{{Axion-like particles
  as mediators for dark matter: beyond freeze-out}},
  \href{https://doi.org/10.1007/JHEP02(2023)141}{\emph{JHEP} {\bfseries 02}
  (2023) 141} [\href{https://arxiv.org/abs/2209.03932}{{\ttfamily
  2209.03932}}].

\bibitem{Armando:2023zwz}
G.~Armando, P.~Panci, J.~Weiss and R.~Ziegler, \emph{{Leptonic ALP portal to
  the dark sector}},
  \href{https://doi.org/10.1103/PhysRevD.109.055029}{\emph{Phys. Rev. D}
  {\bfseries 109} (2024) 055029}
  [\href{https://arxiv.org/abs/2310.05827}{{\ttfamily 2310.05827}}].

\bibitem{Buttazzo:2020vfs}
D.~Buttazzo, P.~Panci, D.~Teresi and R.~Ziegler, \emph{{Xenon1T excess from
  electron recoils of non-relativistic Dark Matter}},
  \href{https://doi.org/10.1016/j.physletb.2021.136310}{\emph{Phys. Lett. B}
  {\bfseries 817} (2021) 136310}
  [\href{https://arxiv.org/abs/2011.08919}{{\ttfamily 2011.08919}}].

\bibitem{Bandyopadhyay:2021wbb}
T.~Bandyopadhyay, S.~Ghosh and T.S.~Roy, \emph{{ALP-Pions generalized}},
  \href{https://doi.org/10.1103/PhysRevD.105.115039}{\emph{Phys. Rev. D}
  {\bfseries 105} (2022) 115039}
  [\href{https://arxiv.org/abs/2112.13147}{{\ttfamily 2112.13147}}].

\bibitem{Michel50}
\emph{{L.~Michel, Proc. Phys. Soc. \textbf{A63}, 514 (1950); C.~Bouchiat and
  L.~Michel, Phys. Rev. \textbf{106}, 170 (1957); T.~Kinoshita and A.~Sirlin,
  Phys. Rev. \textbf{107}, 593 (1957); T.~Kinoshita and A.~Sirlin, Phys. Rev.
  \textbf{108}, 844 (1957)}}, .

\bibitem{TWIST_Apparatus:2005}
R.H.~{\it et al.}, \emph{{TWIST Apparatus}}, {\emph{Nucl. Instrum. Methods A
  \textbf{548}, 306} (2005) }.

\bibitem{TWIST:2011egd}
{\scshape TWIST} collaboration, \emph{{Precise measurement of parity violation
  in polarized muon decay}},
  \href{https://doi.org/10.1103/PhysRevD.84.032005}{\emph{Phys. Rev. D}
  {\bfseries 84} (2011) 032005}
  [\href{https://arxiv.org/abs/1104.3632}{{\ttfamily 1104.3632}}].

\bibitem{TWIST:2011aa}
{\scshape TWIST} collaboration, \emph{{Precision muon decay measurements and
  improved constraints on the weak interaction}},
  \href{https://doi.org/10.1103/PhysRevD.85.092013}{\emph{Phys. Rev. D}
  {\bfseries 85} (2012) 092013}
  [\href{https://arxiv.org/abs/1112.3606}{{\ttfamily 1112.3606}}].

\bibitem{TWIST:2008myj}
{\scshape TWIST} collaboration, \emph{{A Precision Measurement of the Muon
  Decay Parameters rho and delta}},
  \href{https://doi.org/10.1103/PhysRevD.78.032010}{\emph{Phys. Rev. D}
  {\bfseries 78} (2008) 032010}
  [\href{https://arxiv.org/abs/0807.1125}{{\ttfamily 0807.1125}}].

\bibitem{TWIST:2004lzd}
{\scshape TWIST} collaboration, \emph{{Measurement of the muon decay parameter
  delta}}, \href{https://doi.org/10.1103/PhysRevD.71.071101}{\emph{Phys. Rev.
  D} {\bfseries 71} (2005) 071101}
  [\href{https://arxiv.org/abs/hep-ex/0410045}{{\ttfamily hep-ex/0410045}}].

\bibitem{TWIST:2004hce}
{\scshape TWIST} collaboration, \emph{{Measurement of the Michel parameter rho
  in muon decay}},
  \href{https://doi.org/10.1103/PhysRevLett.94.101805}{\emph{Phys. Rev. Lett.}
  {\bfseries 94} (2005) 101805}
  [\href{https://arxiv.org/abs/hep-ex/0409063}{{\ttfamily hep-ex/0409063}}].

\bibitem{Charm:1985}
F.B.~et~al., \emph{{CHARM Collaboration, Search for Axion Like Particle
  Production in 400 - GeV Proton - Copper Interactions}}, {\emph{Phys. Lett. B
  157 (1985) 458–462} (1985) }.

\bibitem{Bjorkeroth:2018dzu}
F.~Bj\"orkeroth, E.J.~Chun and S.F.~King, \emph{{Flavourful Axion
  Phenomenology}}, \href{https://doi.org/10.1007/JHEP08(2018)117}{\emph{JHEP}
  {\bfseries 08} (2018) 117}
  [\href{https://arxiv.org/abs/1806.00660}{{\ttfamily 1806.00660}}].

\bibitem{PIONEER:2022yag}
{\scshape PIONEER} collaboration, \emph{{PIONEER: Studies of Rare Pion
  Decays}},  \href{https://arxiv.org/abs/2203.01981}{{\ttfamily 2203.01981}}.

\bibitem{Poblaguev:2002ug}
A.A.~Poblaguev et~al., \emph{{Experimental study of the radiative decays $K^+
  \rightarrow \mu^+ \nu e^+ e^-$ and $K^+ \rightarrow e^+ \nu e^+ e^-$}},
  \href{https://doi.org/10.1103/PhysRevLett.89.061803}{\emph{Phys. Rev. Lett.}
  {\bfseries 89} (2002) 061803}
  [\href{https://arxiv.org/abs/hep-ex/0204006}{{\ttfamily hep-ex/0204006}}].

\bibitem{Calibbi:2020jvd}
L.~Calibbi, D.~Redigolo, R.~Ziegler and J.~Zupan, \emph{{Looking forward to
  lepton-flavor-violating ALPs}},
  \href{https://doi.org/10.1007/JHEP09(2021)173}{\emph{JHEP} {\bfseries 09}
  (2021) 173} [\href{https://arxiv.org/abs/2006.04795}{{\ttfamily
  2006.04795}}].

\bibitem{Bilger:1998rp}
R.~Bilger, K.~Foehl, H.~Clement, M.~Croni, A.~Erhardt, R.~Meier et~al.,
  \emph{{Search for exotic muon decays}},
  \href{https://doi.org/10.1016/S0370-2693(98)01507-X}{\emph{Phys. Lett. B}
  {\bfseries 446} (1999) 363}
  [\href{https://arxiv.org/abs/hep-ph/9811333}{{\ttfamily hep-ph/9811333}}].

\bibitem{TWIST:2014ymv}
{\scshape TWIST} collaboration, \emph{{Search for two body muon decay
  signals}}, \href{https://doi.org/10.1103/PhysRevD.91.052020}{\emph{Phys. Rev.
  D} {\bfseries 91} (2015) 052020}
  [\href{https://arxiv.org/abs/1409.0638}{{\ttfamily 1409.0638}}].

\bibitem{PIENU:2021clt}
{\scshape PIENU} collaboration, \emph{{Search for three body pion decays
  ${\pi}^+{\to}l^+{\nu}X$}},
  \href{https://doi.org/10.1103/PhysRevD.103.052006}{\emph{Phys. Rev. D}
  {\bfseries 103} (2021) 052006}
  [\href{https://arxiv.org/abs/2101.07381}{{\ttfamily 2101.07381}}].

\bibitem{PIENU:2020loi}
{\scshape PIENU} collaboration, \emph{{Improved search for two body muon decay
  ${\mu}^+{\rightarrow}e^+X_H$}},
  \href{https://doi.org/10.1103/PhysRevD.101.052014}{\emph{Phys. Rev. D}
  {\bfseries 101} (2020) 052014}
  [\href{https://arxiv.org/abs/2002.09170}{{\ttfamily 2002.09170}}].

\bibitem{Nelson:2011sf}
A.E.~Nelson and J.~Scholtz, \emph{{Dark Light, Dark Matter and the Misalignment
  Mechanism}}, \href{https://doi.org/10.1103/PhysRevD.84.103501}{\emph{Phys.
  Rev. D} {\bfseries 84} (2011) 103501}
  [\href{https://arxiv.org/abs/1105.2812}{{\ttfamily 1105.2812}}].

\bibitem{Arias:2012az}
P.~Arias, D.~Cadamuro, M.~Goodsell, J.~Jaeckel, J.~Redondo and A.~Ringwald,
  \emph{{WISPy Cold Dark Matter}},
  \href{https://doi.org/10.1088/1475-7516/2012/06/013}{\emph{JCAP} {\bfseries
  06} (2012) 013} [\href{https://arxiv.org/abs/1201.5902}{{\ttfamily
  1201.5902}}].

\bibitem{MEGII:2021fah}
{\scshape MEG II} collaboration, \emph{{The Search for \ensuremath{\mu}+
  \textrightarrow{} e+\ensuremath{\gamma} with 10\textendash{}14 Sensitivity:
  The Upgrade of the MEG Experiment}},
  \href{https://doi.org/10.3390/sym13091591}{\emph{Symmetry} {\bfseries 13}
  (2021) 1591} [\href{https://arxiv.org/abs/2107.10767}{{\ttfamily
  2107.10767}}].

\bibitem{Meucci:2022qbh}
{\scshape MEG II} collaboration, \emph{{MEG II experiment status and
  prospect}}, \href{https://doi.org/10.22323/1.402.0120}{\emph{PoS} {\bfseries
  NuFact2021} (2022) 120} [\href{https://arxiv.org/abs/2201.08200}{{\ttfamily
  2201.08200}}].

\bibitem{Bandyopadhyay:2022bgx}
T.~Bandyopadhyay, A.~Budhraja, S.~Mukherjee and T.S.~Roy, \emph{{A twisted tale
  of the transverse-mass tail}},
  \href{https://doi.org/10.1007/JHEP08(2023)135}{\emph{JHEP} {\bfseries 08}
  (2023) 135} [\href{https://arxiv.org/abs/2212.02534}{{\ttfamily
  2212.02534}}].

\bibitem{PhysRevLett.60.1793}
G.~Raffelt and D.~Seckel, \emph{{Bounds on exotic-particle interactions from
  SN1987A}}, \href{https://doi.org/10.1103/PhysRevLett.60.1793}{\emph{Phys.
  Rev. Lett.} {\bfseries 60} (1988) 1793}.

\bibitem{Patt:2006fw}
B.~Patt and F.~Wilczek, \emph{{Higgs-field portal into hidden sectors}},
  \href{https://arxiv.org/abs/hep-ph/0605188}{{\ttfamily hep-ph/0605188}}.

\bibitem{Arcadi:2019lka}
G.~Arcadi, A.~Djouadi and M.~Raidal, \emph{{Dark Matter through the Higgs
  portal}}, \href{https://doi.org/10.1016/j.physrep.2019.11.003}{\emph{Phys.
  Rept.} {\bfseries 842} (2020) 1}
  [\href{https://arxiv.org/abs/1903.03616}{{\ttfamily 1903.03616}}].

\bibitem{Arcadi:2021mag}
G.~Arcadi, A.~Djouadi and M.~Kado, \emph{{The Higgs-portal for dark matter:
  effective field theories versus concrete realizations}},
  \href{https://doi.org/10.1140/epjc/s10052-021-09411-2}{\emph{Eur. Phys. J. C}
  {\bfseries 81} (2021) 653}
  [\href{https://arxiv.org/abs/2101.02507}{{\ttfamily 2101.02507}}].

\end{thebibliography}\endgroup

\end{document}